\documentclass{article}

\usepackage{times}
\usepackage{color}
\usepackage{why3lang}
\usepackage{mycamllang}
\usepackage{isabellelst}
\usepackage{lstcoq}
\definecolor{dkblue}{rgb}{0,0.1,0.5} 
\definecolor{lightblue}{rgb}{0,0.5,0.5} 
\definecolor{dkgreen}{rgb}{0,0.4,0} 
\definecolor{dk2green}{rgb}{0.4,0,0} 
\definecolor{dkviolet}{rgb}{0.6,0,0.8}
\usepackage{graphicx}
\usepackage{url}

\newcommand{\prog}[1]{\textit{#1}}
\newcommand{{\Coq}}{{\textsc{Coq}}}
\newif\ifednotes\ednotesfalse
\ednotestrue   
\long\def\ednote#1#2{\ifednotes
  \par\noindent\framebox{\begin{minipage}{0.99\linewidth}\footnotesize #1: #2\end{minipage}}\par
  \else\relax\fi
}

\ednotesfalse

\begin{document}

\title{Formal Proofs of Tarjan's Algorithm in Why3, {\Coq}, and Isabelle}
\author{
  Ran Chen$^1$,
  Cyril Cohen$^2$,
  Jean-Jacques L\'evy$^3$,\\
  Stephan Merz$^4$, and
  Laurent Th\'ery$^2$\\[5pt]
  $^1$ Iscas, Beijing, China\\
  $^2$ Universit\'e C\^ote d'Azur, Inria, Sophia-Antipolis, France\\
  $^3$ Inria, Paris, France\\
  $^4$ Universit\'e de Lorraine, CNRS, Inria, LORIA, Nancy, France
}
\date{October 2018}

\maketitle

\begin{abstract}\noindent
Comparing provers on a formalization of the same problem
is always a valuable exercise. In this paper, we present
the formal proof of correctness
of a non-trivial algorithm from graph theory that was
carried out in three proof assistants: Why3, {\Coq}, and Isabelle.
\end{abstract}

\section{Introduction}

In this paper, we consider Tarjan's algorithm~\cite{Tarjan72} for
discovering the strongly connected components in a directed graph and
present a formal proof of its correctness in three different systems:
Why3, {\Coq} and Isabelle/HOL. The algorithm is treated at an abstract
level with a functional programming style manipulating finite sets,
stacks and mappings, but it respects the linear time behaviour of the
original presentation. It would not be difficult to derive and prove correct
an efficient implementation with imperative
programs and concrete data types such as integers, linked lists and
mutable arrays from our presentation.

To our knowledge this is the first time that the formal correctness proof of a
non-trivial program is carried out in three very different proof assistants:
Why3 is based on a first-order logic with inductive predicates and
automatic provers, {\Coq} on an expressive theory of higher-order logic and
dependent types, and Isabelle/HOL combines higher-order logic with automatic
provers. We do not claim that our proof is the simplest possible one,
and we will discuss the design and implementation of other proofs in
the conclusion, but our proof is indeed elegant and follows Tarjan's
presentation. Crucially for our comparison, the algorithm is defined at the
same level of abstraction in all three systems, and the proof relies on the same
arguments in the three formal systems. Note that a similar
exercise but for a much more elementary proof (the irrationality of
square root of 2) and using many more proof assistants (17) was presented 
in~\cite{seventeenProvers}.

Formal and informal proofs of algorithms about graphs were already
performed
in~\cite{poskitt-vstte-2010,wengener-2002,pottier-dfs-scc-15,Why3GP15,Lammich:2015:FVD:2676724.2693165,thery-15,nipkow-cade-03,Sergey:2015,Raad2016,Hobor:2013,ChenJJL17}.
Some of them are part a larger library, others focus on the
treatment of pointers or about concurrent algorithms.
In particular, Lammich and Neumann~\cite{Lammich:2015:FVD:2676724.2693165} give
a proof of Tarjan's algorithm within their framework for verifying graph
algorithms in Isabelle/HOL.
In our formalization, we are aiming for a simple, direct, and readable proof.

It is not possible to expose here the details of the full proofs in the
three systems, but the interested reader can access and run them on
the Web~\cite{ChenJJLweb17,CohenThery-SCC17,merz:tarjan}. In this paper, we recall the
principles of the algorithm in section~\ref{sec:algorithm}; we
describe the proofs in the three systems in
sections~\ref{sec:why3}, \ref{sec:coq}, and~\ref{sec:isabelle} by
emphasizing the differences induced by the logic which are used; we conclude in
sections~\ref{sec:proof-comments} and~\ref{sec:conclusion} by commenting the
developments and advantages of each proof system.

\section{The algorithm}
\label{sec:algorithm}

The algorithm~\cite{Tarjan72} performs a depth-first search on the set \prog{vertices}
of all vertices in the graph. Every vertex is visited once and is assigned a
serial number of its visit. The algorithm maintains an environment \prog{e}
containing four fields: a stack \prog{e.stack}, a set \prog{e.sccs} of
strongly connected components, a new fresh serial number \prog{e.sn},
and a function \prog{e.num} which records the serial numbers assigned to vertices.
The field \prog{e.stack} contains the visited vertices which are not
part of the components already stored in \prog{e.sccs}. Vertices
are pushed onto the stack in the order of their visit. 

The depth-first search is organized by two mutually recursive
functions \prog{dfs1} and \prog{dfs}. The function \prog{dfs} takes as
argument a set \prog{r} of roots and an environment \prog{e}. It
returns a pair consisting of an integer and the modified
environment. If the set of roots is empty, the returned integer is
$+\infty$.  Otherwise the returned integer is the minimum of the
results of the calls to \prog{dfs1} on non-visited vertices in
\prog{r} and of the serial numbers of the already visited ones.
%

The main procedure \prog{tarjan} initializes the environment with an
empty stack, an empty set of strongly connected components, the fresh
number $0$ and the constant function giving the number $-1$ to each
vertex. The result is the set of components returned by the function
\prog{dfs} called on all vertices in the graph.

\noindent
\begin{small}
\begin{lstlisting}[language=why3]
let rec dfs1 x e =
  let n0 = e.sn in
  let (n1, e1) = dfs (successors x) 
        (add_stack_incr x e) in	  
  if n1 < n0 then (n1, e1) else 
    let (s2, s3) = split x e1.stack in
    (infty(), {stack = s3; 
       sccs = add (elements s2) e1.sccs;
       sn = e1.sn; num = set_infty s2 e1.num}) 

with dfs r e = if is_empty r then (infty(), e) else
  let x = choose r in
  let r' = remove x r in
  let (n1, e1) = if e.num[x] <> -1 
     then (e.num[x], e) else dfs1 x e in
  let (n2, e2) = dfs r' e1 in (min n1 n2, e2)

let tarjan () =
  let e = {stack = Nil; sccs = empty; 
           sn = 0; num = const (-1)} in
  let (_, e') = dfs vertices e in e'.sccs
\end{lstlisting}
\end{small}

The heart of the algorithm is in the body of \prog{dfs1} which visits
a new vertex \prog{x}. The auxiliary function \prog{add\_stack\_incr} updates
the environment by pushing \prog{x} on the stack, assigning it the current fresh
serial number, and incrementing that number in view of future calls. The
function \prog{dfs1} performs a recursive call to \prog{dfs} for the successor
vertices of \prog{x} as roots and the updated environment.
If the returned integer value \prog{n1} is less than the number assigned to
\prog{x}, the function simply returns \prog{n1} and the current environment.
Otherwise, the function declares that a new strongly connected
component has been found, consisting of all vertices that are contained
on top of \prog{x} in the current stack. Therefore the
stack is popped until \prog{x}; the popped vertices are stored as a
new set in \prog{e.sccs}; and their numbers are all set to $+\infty$,
ensuring that they do not interfere with future calculations of min
values. The auxiliary functions \prog{split} and \prog{set\_infty} are used to
carry out these updates.

\begin{small}
\begin{lstlisting}[language=why3]
let add_stack_incr x e = let n = e.sn in
  {stack = Cons x e.stack; sccs = e.sccs; 
      sn = n+1; num = e.num[x <- n]}

let rec set_infty s f = match s with Nil -> f
  | Cons x s' -> (set_infty s' f)[x <- infty()]  end

let rec split x s = match s with Nil -> (Nil, Nil)
  | Cons y s' -> if x = y then (Cons x Nil, s') 
    else let (s1', s2) = split x s' in 
         (Cons y s1', s2)  end
\end{lstlisting}
\end{small}

\begin{figure}[t]
\begin{center}
\includegraphics[width=.6\linewidth]{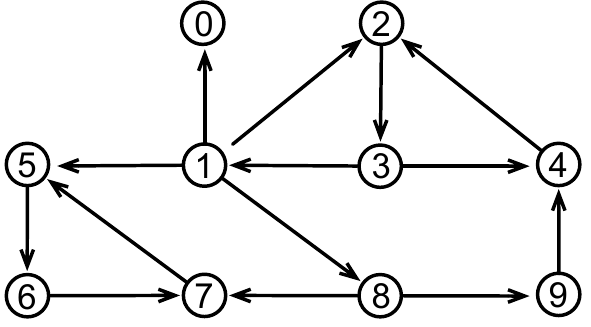}
\end{center}%
\vspace*{-1em}
\caption{An example: the vertices are numbered and
  pushed onto the stack in the order of their visit by the recursive
  function \prog{dfs1}. When the first component $\{0\}$ is
  discovered, vertex $0$ is popped; similarly when the second
  component $\{5, 6, 7\}$ is found, its vertices are popped; finally
  all vertices are popped when the third component $\{1,2,3,4,8,9\}$
  is found. Notice that there is no cross-edge to a vertex with a number
  less than $5$ when the second component is discovered. Similarly in
  the first component, there is no edge to a vertex with a number less
  than $0$. In the third component, there is no edge to a vertex
  less than $1$ since we have set the number of vertex $0$ to
  $+\infty$ when $0$ was popped.}
\label{fig:example1}
\end{figure}

Figure~\ref{fig:example1} illustrates the behavior of the algorithm by an
example. We presented the algorithm as a functional program, using data
structures available in the Why3 standard library~\cite{why3manual0861}. For
lists we have the constructors \prog{Nil} and \prog{Cons}; the function
\prog{elements} returns the set of elements of a list. For finite sets, we have
the empty set \prog{empty}, and the functions \prog{add} to add an element to a
set, \prog{remove} to remove an element from a set, \prog{choose} to pick an
arbitrary element in a (non-empty) set, and \prog{is\_empty} to test for
emptiness. We also use maps with functions \prog{const} denoting the constant
function, \texttt{\_[\_]} to access the value of an element, and
\mbox{\texttt{\_[\_ $\leftarrow$ \_]}} for creating a map obtained from an
existing map by setting an element to a given value. We also define an abstract
type \prog{vertex} for vertices and a constant \prog{vertices} for the finite
set of all vertices in the graph. The type \prog{env} of environments is a
record with the four fields \prog{stack}, \prog{sccs}, \prog{sn} and \prog{num}
as described above.

\begin{small}
\begin{lstlisting}[language=why3]
type vertex
constant vertices: set vertex
function successors vertex : set vertex
type env = {stack: list vertex; 
            sccs: set (set vertex);
            sn: int; num: map vertex int}
\end{lstlisting}
\end{small}

For a correspondence between our presentation and the imperative
programs used in standard textbooks, the reader is referred
to~\cite{ChenJJL17}. The present version can be directly translated
into {\Coq} or Isabelle functions, and it respects the linear running
time behaviour of the algorithm, since vertices could be easily
implemented by integers, $+\infty$ by the cardinal of \prog{vertices},
finite sets by lists of integers and mappings by mutable arrays (see
for instance~\cite{ChenJJLweb17}).

\begin{figure}[b]
\begin{center}
\begin{tabular}{l@{\hspace{.15\linewidth}}l}
\begin{tabular}[c]{l}
\includegraphics[width=.4\linewidth]{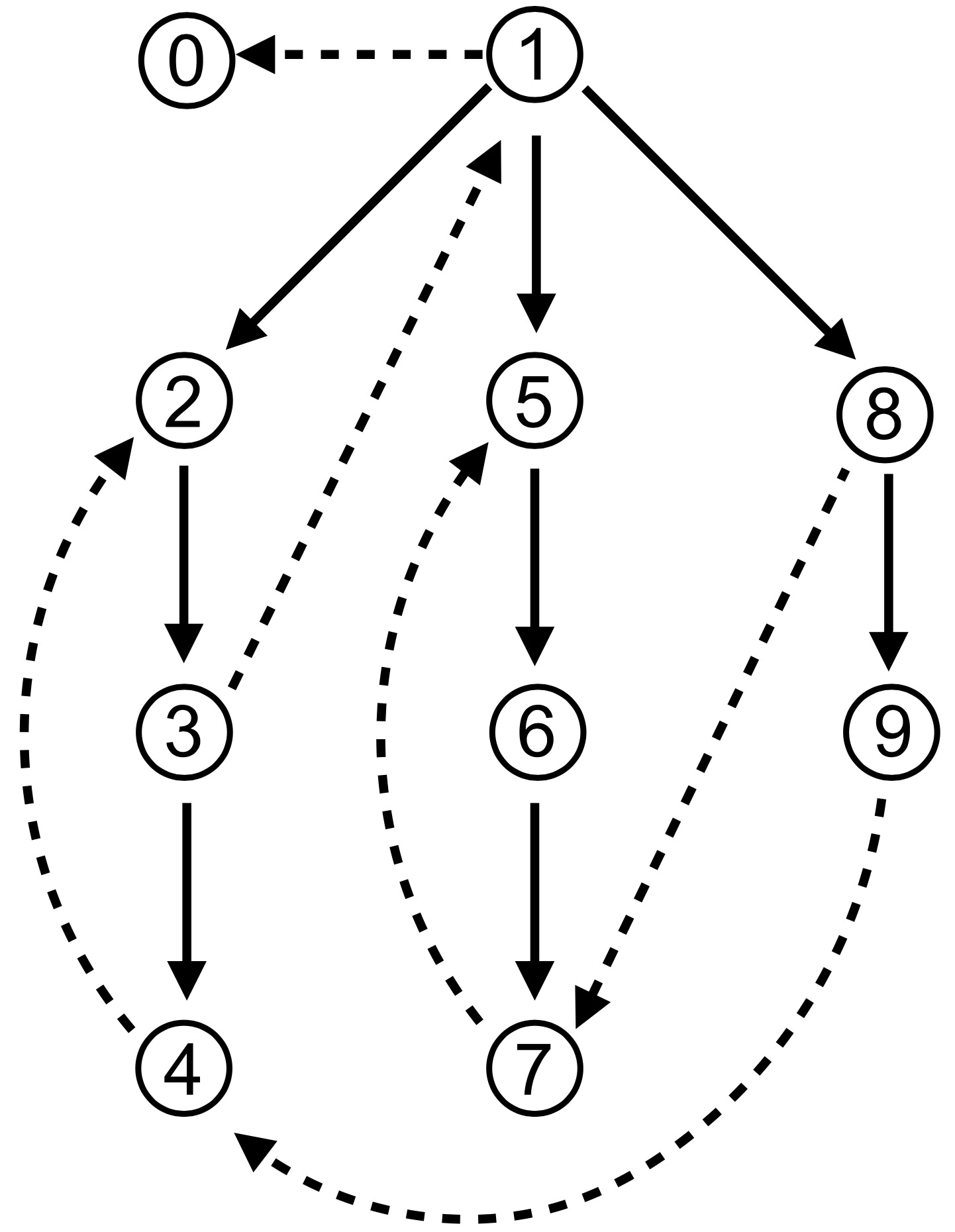}
\end{tabular}
& %
\begin{small}
$\begin{array}[c]{l@{\hspace{.07\linewidth}}l}
0 & 0 \\
1 & 1 \\
2 & 1 \\
3 & 1 \\
4 & 2 \\
5 & 5 \\
6 & 5 \\
7 & 5 \\
8 & 4 \\
9 & 4 \\[4ex]
\end{array}$
\end{small}
\end{tabular}
\end{center}%
\vspace*{-1em}
\caption{Spanning forest and the $\mathop{\prog{LOWLINK}}$ function.}
\label{fig:spanning-forest}
\end{figure}

Like many algorithms on graphs, Tarjan's algorithm is not easy to understand and
even looks a bit magical. In the
original presentation, the integer value returned by the function
\prog{dfs1}\/ is given by the following formula when called on vertex
\prog{x}.
\newcommand{\treeedge}{\Longrightarrow}
\newcommand{\treepath}{\stackrel{*}{\Longrightarrow}}
\newcommand{\xedge}{\hookrightarrow}
\newcommand{\LOWLINK}{\mathop{\prog{LOWLINK}}}
$$\begin{array}{l}
\LOWLINK(x) = \min\{\prog{num}[y] \;\mid\;  x \treepath z \xedge y \\ 
    \hspace{4ex} \wedge\mbox{ $x$ and $y$ are in the same connected component} \}
\end{array}
$$
This expression is evaluated on the spanning tree (forest) corresponding to one
run of \prog{dfs}. The relation $x \treeedge z$ means that $z$ is a
son of $x$ in the spanning tree, the relation $\treepath$ is its
transitive and reflexive closure, and $z \xedge y$ means that there is
a cross-edge between \prog{z} and \prog{y} in the spanning tree. In
figure~\ref{fig:spanning-forest}, $\treeedge$ is drawn in thick lines
and $\xedge$ in dotted lines; a table of the values of the
\prog{LOWLINK} function is also shown. Thus the integer value returned
by \prog{dfs1} is the minimum of the numbers of vertices in the same
connected component accessible by just one cross-edge by all
descendants of \prog{x} visited in the recursive calls. If none,
$+\infty$ is returned (here is a slight simplification w.r.t.\ the original
algorithm). Notice that the result may be the number of a vertex which is
not an ancestor of \prog{x} in the spanning tree. Take for instance,
vertices $8$ or $9$ in figure~\ref{fig:spanning-forest}.

The algorithm relies on the existence of a base with a minimal serial
number for each connected component, the members of which are among its
descendants in the spanning tree. The reason is that a cross-edge
reaches from $x$ either an ancestor of $x$, or a descendant of a
grandson in the spanning tree, or a cousin to the left of
$x$. Intuitively, cross-edges never go right in the spanning
tree. Therefore these bases are organized as a Christmas tree, and
each connected component is a prefix of one sub-tree of which the root
is its base.

Thus for each environment \prog{e} in the algorithm, the working stack
\prog{e.stack} corresponds to a cut of the spanning tree where
connected components to its left are pruned and stored in
\prog{e.sccs}. In this stack, any vertex can reach any vertex higher
in the stack. And if a vertex is a base of a connected component, no
cross-edge can reach some vertex lower than this base in the stack,
otherwise that last vertex would be in the same connected component
with a strictly lower serial number.

We therefore have to organize the proofs of the algorithm around these
arguments. To maintain these invariants we will distinguish, as is common for
depth-first search algorithms, three sets of vertices: white vertices are the
non-visited ones, black vertices are those that are already fully visited, and gray
vertices are those that are still being visited. Clearly, these sets are disjoint
and white vertices can be considered as forming the complement in
\prog{vertices} of the union of the gray and black ones.

The previously mentioned invariant properties can now be expressed for
vertices in the stack: no such vertex is white, any 
vertex can
reach all vertices higher in the stack, any vertex can reach some gray
vertex lower in the stack.  Moreover, vertices in the stack respect the
numbering order, i.e.\ a vertex \prog{x} is lower than \prog{y} in the
stack if and only if the number assigned to \prog{x} is strictly less
than the number assigned to \prog{y}.

\section{The proof in Why3}
\label{sec:why3}

The Why3 system comprises the language WhyML for writing programs and a
many sorted first-order logic with inductive data types and inductive
predicates to express the logical assertions. The system generates proof
obligations w.r.t.\ the assertions, pre- and post-conditions and lemmas
inserted in the WhyML program. The system is interfaced with
off-the-shelf automatic provers (we mainly use Alt-Ergo, CVC, E-prover
and Z3) and also interactive proof assistants such as {\Coq} or Isabelle.

There are numerous libraries that can be used in the Why3 library,
for integer arithmetic, polymorphic lists, finite sets and mappings,
etc. There is also a small theory for paths in graphs. Here we define
graphs, paths and strongly connected components as follows.
\begin{small}
\begin{lstlisting}[language=why3]
axiom successors_vertices: forall x. mem x vertices -> 
  subset (successors x) vertices
predicate edge (x y: vertex) = 
  mem x vertices /\ mem y (successors x)
inductive path vertex (list vertex) vertex =
  | Path_empty: forall x: vertex. path x Nil x
  | Path_cons: forall x y z: vertex, l: list vertex.
      edge x y -> path y l z -> path x (Cons x l) z

predicate reachable (x y: vertex) = exists l. path x l y
predicate in_same_scc (x y: vertex)  =  
  reachable x y /\ reachable y x
predicate is_subscc (s: set vertex) =  
  forall x y. mem x s -> mem y s -> in_same_scc x y
predicate is_scc (s: set vertex) = not is_empty s 
  /\ is_subscc s
  /\ (forall s'. subset s s' -> is_subscc s' -> s == s')
\end{lstlisting}
\end{small}
where \prog{mem} and \prog{subset} denote membership and the subset
relation for finite sets.

We add two ghost fields in environments for the black and gray sets of
vertices. These fields are used in the proofs but not used in the
calculation of the connected components, which is checked by the
type-checker of the language.

\begin{small}
\begin{lstlisting}[language=why3]
type env = {ghost black: set vertex; 
    ghost gray: set vertex;
    stack: list vertex; sccs: set (set vertex);
    sn: int; num: map vertex int}
\end{lstlisting}
\end{small}
The functions now become:
\begin{small}
\begin{lstlisting}[language=why3]
let rec dfs1 x e =
  let n0 = e.sn in
  let (n1, e1) = dfs (successors x) 
        (add_stack_incr x e) in	  
  if n1 < n0 then (n1, add_black x e1) else 
    let (s2, s3) = split x e1.stack in
    (infty(), {stack = s3; 
       black = add x e1.black; gray = e.gray;
       sccs = add (elements s2) e1.sccs;
       sn = e1.sn; num = set_infty s2 e1.num}) 
with dfs r e = ... (* unmodified *)
let tarjan () =
  let e = {black = empty; gray = empty; 
    stack = Nil; sccs = empty; sn = 0; 
    num = const (-1)} in
  let (_, e') = dfs vertices e in e'.sccs
\end{lstlisting}
\end{small}
with a new function \prog{add\_black} turning a vertex from gray to black and
the modified \prog{add\_stack\_incr} adding a new gray vertex with a
fresh serial number to the current stack.
\begin{small}
\begin{lstlisting}[language=why3]
let add_stack_incr x e =
  let n = e.sn in
  {black = e.black; gray = add x e.gray; 
   stack = Cons x e.stack; sccs = e.sccs; 
   sn = n+1; num = e.num[x <-n]}
let add_black x e =
 {black = add x e.black; gray = remove x e.gray; 
  stack = e.stack; sccs = e.sccs; 
  sn = e.sn; num = e.num}
\end{lstlisting}
\end{small}
\newcommand{\invI}{$(\mathcal{I})$}
\noindent
The main invariant \invI\ of our program states that the environment
is well-formed:
\begin{small}
\begin{lstlisting}[language=why3]
predicate wf_env (e: env) = 
  let {stack = s; black = b; gray = g} = e in
  wf_color e /\ wf_num e /\
  simplelist s /\ no_black_to_white b g /\ 
  (forall x y. lmem x s -> lmem y s -> 
     e.num[x] <= e.num[y] -> reachable x y) /\
  (forall y. lmem y s -> exists x. mem x g /\ 
     e.num[x] <= e.num[y] /\ reachable y x) /\
  (forall cc. mem cc e.sccs <-> 
              subset cc e.black /\ is_scc cc)
\end{lstlisting}
\end{small}
where \prog{lmem} stands for membership in a list. 
The well-formed\-ness property is the conjunction of seven clauses. The two
first clauses express quite elementary conditions about the colored sets of vertices
and the numbering function. We do not express them formally here
(see~\cite{ChenJJL17,ChenJJLweb17} for a detailed description). The
third clause states that there are no repetitions in the stack, and
the fourth that there is no edge from a black vertex to a white
vertex. The next two clauses formally express the property already
stated above: any vertex in the stack reaches all higher vertices and
any vertex in the stack can reach a lower gray vertex. The last clause states
that the \prog{sccs} field is the set of all connected
components all of whose vertices are black. 
Since at the end of the \prog{tarjan} function,
all vertices are black, the \prog{sccs} field will contain exactly the set of all
strongly connected components.

Our functions \prog{dfs1} and \prog{dfs} modify the environment in a
monotonic way. Namely they augment the set of the fully visited
vertices (the black ones); they keep invariant the set of the ones
currently under visit (the gray set); they increase the stack with new
black vertices; they also discover new connected components and they
keep invariant the serial numbers of vertices in the stack,
\begin{small}
\begin{lstlisting}[language=why3]
predicate subenv (e e': env) =
  subset e.black e'.black /\ e.gray == e'.gray  
  /\ (exists s. e'.stack = s ++ e.stack /\ 
         subset (elements s) e'.black) 
  /\ subset e.sccs e'.sccs 
  /\ (forall x. lmem x e.stack -> e.num[x] = e'.num[x])
\end{lstlisting}
\end{small}
Once these invariants are expressed, it remains to locate them in the
program text and to add assertions which help to prove them. The
pre-conditions of \prog{dfs1} are quite natural: the vertex \prog{x}
must be a white vertex of the graph, and it must be
reachable from all gray vertices. Moreover invariant \invI\ must
hold. The post-conditions of \prog{dfs1} are of three kinds. Firstly
\invI\ and the monotony property \prog{subenv} hold in the resulting
environment. Vertex \prog{x} is black at the end of \prog{dfs1}. Finally
we express properties of the integer value \prog{n} returned by this
function which should be $\LOWLINK(x)$ as announced previously. Notice
that we do not know yet the connected component of \prog{x}, but the
definition of $\LOWLINK$ still works thanks to the numbering with
$+\infty$ of the visited vertices not in its component. In this
proof, we give three implicit properties for characterizing the resulting
\prog{n} value. First, the returned value is never higher than the
number of \prog{x} in the final environment. Secondly, the returned value
is either $+\infty$ or the number of a vertex in the stack reachable
from \prog{x}. Finally, if there is an edge from a vertex \prog{y'} in
the new part of the stack to a vertex \prog{y} in its old part, the
resulting value \prog{n} must be lower than the number of \prog{y}.
\begin{small}
\begin{lstlisting}[language=why3]
let rec dfs1 x e  =
(* pre-condition *)
requires {mem x vertices}
requires {forall y. mem y e.gray -> reachable y x}
requires {not mem x (union e.black e.gray)}
requires {wf_env e} (* I *)
(* post-condition *)  
returns {(_, e') -> wf_env e' /\ subenv e e'} 
returns {(_, e') -> mem x e'.black} 
returns {(n, e') -> n <= e'.num[x]}
returns {(n, e') -> n = infty() \/ num_of_reachable n x e'}
returns {(n, e') -> forall y. xedge_to e'.stack e.stack y 
                    -> n <= e'.num[y]}
\end{lstlisting}
\end{small}
where the auxiliary predicates used in these post-conditions are
formally defined in the following way.
\begin{small}
\begin{lstlisting}[language=why3]
predicate num_of_reachable (n: int) (x: vertex) 
  (e: env) = exists y. lmem y e.stack /\ n = e.num[y] /\ 
     reachable x y
predicate xedge_to (s1 s3: list vertex) 
  (y: vertex) = (exists s2. s1 = s2 ++ s3 /\ 
    exists y'. lmem y' s2 /\ edge y' y) /\ lmem y s3
\end{lstlisting}
\end{small}
Notice that when the integer result \prog{n} of \prog{dfs1} is infinite, the
number of \prog{x} must also be infinite, meaning that its connected
component has been found. Also notice that the definition of
\prog{xedge\_to} fits the definition of $\LOWLINK$ when the cross edge
ends at a vertex residing in the stack before the call of
\prog{dfs1}.
The pre- and post-conditions for the function \prog{dfs} are quite
similar up to a generalization to sets of vertices which are the
roots of the algorithm (see~\cite{ChenJJLweb17}). 

We now add seven assertions in the body of the \prog{dfs1} function to
help the automatic provers. In contrast, the function \prog{dfs} needs
no extra assertions in its body.  In \prog{dfs1}, when the number
\prog{n0} of \prog{x} is strictly greater than the number resulting
from the call to its successors, the first assertion states that
vertex \prog{x} can reach a strictly lower vertex in the current stack
and the second assertion states that a lower gray vertex is reachable
and that thus the connected component of \prog{x} is not fully black
at end of \prog{dfs1}.  That key assertion is proved from the first
one by transitivity of reachability.
(We understand here why the algorithm only takes care of a single
cross-edge: for any visited vertex \prog{x}, the spanning tree 
must contain at least one back-edge to a strict ancestor of
\prog{x} when $\prog{n1} < \prog{n0}\,$.)
The next four assertions show that the
connected component \prog{(elements s2)} of \prog{x} is on top of
\prog{x} in the stack when $\prog{n1} \geq \prog{n0}$, and that then
\prog{x} is the base of that connected component. The seventh
assertion helps proving that the coloring constraint is preserved at the end of
\prog{dfs1}.
\begin{small}
\begin{lstlisting}[language=why3]
let n0 = e.sn in
let (n1, e1) =
  dfs (successors x) (add_stack_incr x e) in	  
if n1 < n0 then begin
  assert {exists y. y <> x /\ precedes x y e1.stack /\ 
     reachable x y}; 
  assert {exists y. y <> x /\ mem y e1.gray /\ 
     e1.num[y] < e1.num[x] /\ in_same_scc x y};
  (n1, add_black x e1) end
else
  let (s2, s3) = split x e1.stack in
  assert {is_last x s2 /\ s3 = e.stack /\ 
     subset (elements s2) (add x e1.black)};
  assert {is_subscc (elements s2)};
  assert {forall y. in_same_scc y x -> lmem y s2};  
  assert {is_scc (elements s2)};
  assert {inter e.gray (elements s2) == empty}; 
  (infty(), {black = add x e1.black; gray = e.gray;
     stack = s3; sccs = add (elements s2) e1.sccs;
     sn = e1.sn; num = set_infty s2 e1.num}) 
\end{lstlisting}
\end{small}
where \prog{inter} is set intersection, and \prog{precedes} and
\prog{is\_last} are two auxiliary predicates defined below.
\begin{small}
\begin{lstlisting}[language=why3]
predicate is_last (x: 'a) (s: list 'a) =
  exists s'. s = s' ++ Cons x Nil 
predicate precedes (x y: 'a) (s: list 'a) =
  exists s1 s2. s = s1 ++ (Cons x s2) /\ lmem y (Cons x s2)
\end{lstlisting}
\end{small}
All proofs are discovered by the automatic provers except for two
proofs carried out interactively in {\Coq}. One is the proof of the black
extension of the stack in case $\prog{n1} < \prog{n0}$. The provers
could not work with the existential quantifier, although the {\Coq} proof
is quite short. The second {\Coq} proof is the fifth assertion in the
body of \prog{dfs1}, which asserts that any \prog{y} in the connected
component of \prog{x} belongs to \prog{s2}. It is a maximality
assertion which states that the set \prog{(elements s2)} is a complete
connected component. The proof of that assertion is by
contradiction. If \prog{y} is not in \prog{s2}, there must be an edge
from \prog{x'} in \prog{s2} to some \prog{y'} not in \prog{s2} such
that \prog{x} reaches \prog{x'} and \prog{y'} reaches \prog{y}. There
are three cases, depending on the position of \prog{y'}. Case 1 is
when \prog{y'} is in \prog{sccs}: this is not possible since \prog{x}
would then be in \prog{sccs} which contradicts \prog{x} being
gray. Case 2 is when \prog{y'} is an element of \prog{s3}: the serial
number of \prog{y'} is strictly less than the one of \prog{x} which is
\prog{n0}. If $\prog{x'} \neq\prog{x}$, the cross-edge from \prog{x'}
to \prog{y'} contradicts $\prog{n1} \geq \prog{n0}$ (post-condition
5); if $\prog{x'} = \prog{x}$, then \prog{y'} is a successor of
\prog{x} and again it contradicts $\prog{n1} \geq \prog{n0}$
(post-condition 3). Case 3 is when \prog{y'} is white, which is
impossible since \prog{x'} is black in \prog{s2}.

\newcommand{\hh}{\hspace*{.5ex}}
\newcommand{\hhseven}{\hh\hh\hh\hh\hh\hh\hh}
\begin{table}[t]
\begin{center}
\begin{footnotesize}
\begin{tabular}{|l|r|r|r|r|r||@{\hh\hh}r|r|}\hline
provers &{\sf Alt-}\hh\hh&{\sf CVC4}&{\sf E-}\hhseven&\hh\hh{\sf Z3}\hh\hh&\#VC&\#PO\\ 
        &{\sf Ergo}\hh&   &{\sf prover}& & &\\ \hline\hline  
49 lemmas        & 1.91   & 26.11 & 3.33 &      & 70 & 49\\
split            &0.09    &  0.16 &      &      & 6  & 6\\
add\_stack\_incr &0.01    &      &      &      & 1  & 1\\
add\_black       &0.02    &      &      &      & 1  & 1 \\
set\_infty       &0.03    &      &      &      & 1  & 1\\
dfs1             &77.89   &150.2& 19.99& 13.67 &79 & 20\\
dfs              &4.71     &3.52 &      & 0.26 & 58 & 25 \\
tarjan           &0.85    &      &      &      &15  & 5 \\\hline\hline
total            &85.51   &179.99 &23.32 & 13.93 & 231 & 108\\
\hline
\end{tabular}
\end{footnotesize}
\end{center}
\caption{Performance results of the provers (in seconds, on a 3.3 GHz Intel Core i5
  processor). Total time is 341.15 seconds. The two last columns contain the
  numbers of verification conditions and proof obligations. Notice that there
  may be several VCs per proof obligation.}
\label{table:results}
\end{table}

The figures of the Why3 proof are listed in
table~\ref{table:results}. Alt-Ergo 2.2 and CVC4 1.5 proved the bulk of the
proof obligations.\footnote{In addition to the results reported in the table, 
  Spass was used to discharge one proof obligation.} %
The proof uses 49
lemmas that were all proved automatically, but with an interactive interface 
providing hints to apply inlining, splitting, or induction strategies. This
includes 13 lemmas on sets, 16 on lists, 5 on lists without repetitions, 3 on
paths, 5 on connected components and 6 very specialized lemmas directly
involved in the proof obligations of the algorithm. Among the lemmas,
a critical one is the lemma \prog{xpath\_xedge} on paths which reduces
a predicate on paths to a predicate on edges. In fact, most of the Why3
proof works on edges which are handled more robustly by the automatic provers
than paths. The two {\Coq} proofs are 16 and 141 lines long (the {\Coq} files of 677 and
721 lines include preambles that are automatically generated during the
translation from Why3 to {\Coq}).
The interested reader is refered to~\cite{ChenJJLweb17} where the
full proof is available.

The proof explained so far does not show that the functions terminate;
we have only shown the partial correctness of the
algorithm. But after adding two lemmas about union and difference for
finite sets, termination is automatically proved by the following
lexicographic ordering on the number of white vertices and roots.
\begin{small}
\begin{lstlisting}[language=why3]
let rec dfs1 x e  =
variant {cardinal (diff vertices 
            (union e.black e.gray)), 0}
with dfs r e = 
variant {cardinal (diff vertices 
            (union e.black e.gray)), 1, cardinal r}
\end{lstlisting}
\end{small}

\section{The proof in {\Coq}}
\label{sec:coq}
{\Coq} is a proof assistant based on type theory. It uses 
the calculus of constructions, a higher order lambda-calculus, to
express formulae and proofs. Some basic notions of graph theory are 
provided by the Mathematical Component Library~\cite{mathcompbook}. 
The formalization in {\Coq} follows closely what has been done in Why3, so 
we mostly highlight differences. It is parameterized by a finite 
type \prog{V} for  the vertices and a function \prog{successors} that 
represents the graph, i.e. \prog{(successors v)} gives all the successors
of the vertex \prog{v} in the graph.

The environment that is passed around in the algorithm
is defined as a record with five fields:
\begin{small}
\begin{lstlisting}[language=Coq]
Record env := Env {
  black : {set V};
  stack : seq V;
  esccs : {set {set V}};
        sn : nat; 
      num : {ffun V -> nat}}.
\end{lstlisting}
\end{small}
Note that with respect to Why3, we have no ghost mechanism available for the \prog{black} field
and we do not hold gray vertices. They are globally defined as the elements of
the stack that are not black. Also, we restrict ourselves to natural numbers, representing
the integer \prog{n} in Why3 by the natural number $\prog{n} + 1$ in {\Coq}.

Our definition of the algorithm is very similar to the one of Why3. The only difference
is the way recursion is handled.
We untangle the mutually recursive function \prog{tarjan} into two separate functions
The first one \prog{dfs1} treats a vertex \prog{x} and 
the second one \prog{dfs} a set of vertices \prog{roots}
in an environment \prog{e}.
\begin{small}
\begin{lstlisting}[language=Coq]
Definition dfs1 dfs x e :=
  let: (m1, e1) := 
    dfs [set y in successors x] (add_stack x e) in
  if m1 < sn e then (m1, add_black x e1)
  else (infty, add_sccs x e1).

Definition dfs dfs1 dfs roots e :=
  if [pick x in roots] isn't Some x then (infty, e)
  else let roots' := roots :\ x in
           let: (m1, e1) :=
               if num e x != 0 then (num e x, e) else dfs1 x e in
           let: (m2, e2) := dfs roots' e1 in (minn m1 m2, e2).
\end{lstlisting}
\end{small}
Then, the two functions are glued together in a recursive function
\prog{tarjan\_rec} where the parameter \prog{n} controls the maximal
recursive height.
\begin{small}
\begin{lstlisting}[language=Coq]
Fixpoint tarjan_rec n :=
  if n is n1.+1 then 
    dfs (dfs1 (tarjan_rec n1)) (tarjan_rec n1)
  else fun r e => (infty, e).

Let N := #|V| * #|V|.+1 + #|V|.
Definition tarjan := sccs (tarjan_rec N setT e0).2.
\end{lstlisting}
\end{small}
If \prog{n} is not zero (i.e. it is a successor of some \prog{n1}),
\prog{tarjan\_rec} calls \prog{dfs} taking care that its parameters can only use 
recursive call to \prog{tarjan\_rec} with a smaller recursive height, here \prog{n1}.
This ensures termination. A dummy value is returned in the case where \prog{n} is zero.
As both \prog{dfs} and \prog{dfs1 } cannot be applied more 
than the number of vertices, the value \prog{N} encodes the lexicographic product of the
two maximal heights. It gives \prog{tarjan\_rec} enough fuel to never encounter the dummy 
value so \prog{tarjan} correctly terminates the computation.
This last statement is of course proved formally later.

The invariants are essentially the same as in the Why3 proof.
There are just packaged in a different way
so we can express more easily intermediate lemmas between the different packages.
We first group together the properties about connectivity
\begin{small}
\begin{lstlisting}[language=Coq]
Record wf_graph e := WfGraph {
  wf_stack_to_stack :
    {in stack e &, forall x y, 
      (num e x <= num e y) -> gconnect x y};
  wf_stack_to_grays : 
    {in stack e, forall y, 
      exists x, [/\ x \in grays e, (num e x <= num e y) & gconnect y x]
}.
\end{lstlisting}
\end{small}
The main invariant then collects all the properties
\begin{small}
\begin{lstlisting}[language=Coq]
Record invariants (e : env) := Invariants {
              inv_wf_color : wf_color e;
                 inv_wf_num  : wf_num e;
              inv_wf_graph : wf_graph e;
  wf_noblack_towhite : noblack_to_white e;
                     inv_sccs  : sccs e = black_gsccs e;
}.
\end{lstlisting}
\end{small}

Pre-conditions are stored in a record and are similar to the ones defined in Why3:
all the gray vertices of \prog{e} are connected to all the elements of \prog{roots}.
and all the invariants hold. 
\begin{small}
\begin{lstlisting}[language=Coq]
Definition access_to e (roots : {set V}) :=
  {in gray e & roots, forall x y, gconnect x y}.

Record pre_dfs (roots : {set V}) (e : env) := PreDfs {
  pre_access_to   : access_to e roots;
  pre_invariants : invariants e
}.
\end{lstlisting}
\end{small}
The post-conditions are expressed slightly differently 
mostly because we take advantage
of the expressivity of big operators~\cite{bigop}. The
\prog{bigcup} operator (typeset as \prog{{\textbackslash}bigcup\_})
is defined in the Mathematical Component
Library and represents indexed union of sets. The
\prog{bigmin} operator (typeset as \prog{{\textbackslash}min\_}) represents
the minimum of a set of natural numbers
(and should be included in future version of the Library).
Defining the minimum of the empty set is a bit problematic since one
would like to preserve the property that the minimum of a subset is never smaller than 
the minimum of the full set. This is why the \prog{bigmin} does not work directly
on sets of natural numbers but on sets of elements of an ordinal type \prog{I$_{n}$}
(the type of all the natural numbers smaller than \prog{n}).
This type has the key property of having a maximal element  \prog{n}. This is the
value given to the minimum of the empty set. In our use case,
as ${\infty}$ is defined as the number of vertices plus one, we simple take
\prog{$n ={\infty}$}.

The post-conditions are then
expressed by a record that states that the invariants hold, the 
next environment is an extension of the old one, 
the new white vertices have been decremented by the vertices that are reachable 
from the roots by white vertices
and finally the returned value \prog{m} is exactly the 
smallest number from all the vertices that have lost their white color.
\begin{small}
\begin{lstlisting}[language=Coq]
Record post_dfs (roots : {set V}) (e e' : env) (m : nat) := PostDfs {
  post_invariants : invariants e';
  post_subenv         : subenv e e';
  post_whites         : 
    whites e' = white e :\: \bigcup_(x in roots) wreach e x;
  post_num               : 
    m = \min_(y in \bigcup_(x in roots) wreach e x)
                    @inord infty (num e' y);
}.
\end{lstlisting}
\end{small}
Note that we have defined the predicate \prog{wreach} to express the reachability
through white vertices and we are using the explicit cast \prog{inord} to turn
a number associated to a vertex into an element of \prog{I$_{\infty}$}.

Now we can state the correctness of \prog{dfs} and
\prog{dfs1}
\begin{small}
\begin{lstlisting}[language=Coq]
Definition dfs_correct
      (dfs : {set V} -> env -> nat * env) roots e :=
  pre_dfs roots e ->
  let (m, e') := dfs roots e in post_dfs roots e e' m.
Definition dfs1_correct
      (dfs1 : V -> env -> nat * env) x e :=
  (x \in white e) -> pre_dfs [set x] e ->
  let (m, e') := dfs1 x e in post_dfs [set x] e e' m.
\end{lstlisting}
\end{small}
where \prog{[set x]} represents the set whose only element is \prog{x}.
The two central theorems to prove are then
\begin{small}
\begin{lstlisting}[language=Coq]
Lemma dfs_is_correct dfs1 dfsrec (roots : {set V}) e :
  (forall x, x \in roots -> dfs1_correct dfs1 x e) ->
  (forall x, x \in roots -> forall e1, white e1 \subset white e ->
         dfs_correct dfsrec (roots :\ x) e1) ->
  dfs_correct (dfs dfs1 dfsrec) roots e. 
Lemma dfs1_is_correct dfs (x : V) e :
  (dfs_correct dfs [set y | edge x y] (add_stack x e)) ->
  dfs1_correct (dfs1 dfs) x e.
\end{lstlisting}
\end{small}
They simply state that the results of \prog{dfs} and \prog{dfs1}
are correct if their respective recursive calls are correct.
The proof of the first lemma is straightforward since \prog{dfs} simply
iterates on a list. It is mostly some book-keeping between what is known and
what needs to be proved. This is done in about 70 lines. The second
one is more intricate and requires 328 lines. Gluing these two theorems
together and proving termination gives us an extra 20 lines to prove
the theorem
\begin{small}
\begin{lstlisting}[language=Coq]
Theorem tarjan_rec_terminates n roots e :
  n >= #|white e| * #|V|.+1 + #|roots| ->
  dfs_correct (tarjan_rec n) roots e.
\end{lstlisting}
\end{small}
From this last theorem the correctness of \prog{tarjan} follows directly.

Some quantitative information should be added. The {\Coq} contribution is composed
of three files. The \prog{bigmin} file defines the \prog{bigmin} operator
and is 160 lines long. The \prog{extra} file defines some notions of graph theory
that were not available in the current Mathematical Component Library.
For example, it is where conditional reachability is defined.
This file is 350 lines long. The main file is \prog{tarjan\_num} and is 
1185 lines long. It is compiled in 10 seconds  with a memory footprint of 900 Mb 
(half of which is resident) on a Intel\textsuperscript{\textregistered} i7 2.60GHz 
quad-core laptop running Linux.
The proofs are performed in the \textsc{SSReflect} proof language~\cite{ssreflect} with very little automation.
The proof script is mostly procedural, alternating book-keeping tactics
({\prog{move}) with transformational ones
(mostly \prog{rewrite} and \prog{apply}), but often
intermediate steps are explicitly declared with the \prog{have} tactic.
There are more than a hundred of such intermediate steps in the 700 lines
of proof of the file \prog{tarjan\_num}. 
Table~\ref{table:coqline}
gives the distribution of the numbers of lines of these proofs.
\begin{table}[t]
\begin{center}
\begin{footnotesize}
\begin{tabular}{|c|c|c|c|c|c|c|}\hline
                 $l = 1$ & $l \le 10$ & $l \le 20$ & $l \le 30$ & 
                 $l = 35$ & $ l = 70 $ & $l = 328$ \cr \hline
                 37 & 25 & 5 & 3 & 1 & 1 & 1 \cr
                 \hline
\end{tabular}
\end{footnotesize}
\end{center}
\caption{Sizes (numbers $l$ of lines) of the 73 proofs in the file \prog{tarjan\_num}.}
\label{table:coqline}
\end{table}
Most of them are one-liners and the only complicated proof is the one corresponding
to the lemma \prog{dfs1\_is\_correct}.

\section{The proof in Isabelle/HOL}
\label{sec:isabelle}

Isabelle/HOL~\cite{nipkow:isabelle} is the encoding of simply typed higher-order
logic in the logical framework Isabelle~\cite{paulson:isabelle}. Unlike Why3, it
is not primarily intended as an environment for program verification and does
not contain specific syntax for stating pre- and post-conditions or intermediate
assertions in function definitions. Logics and formalisms for program
verification have been developed within Isabelle/HOL (e.g.,
\cite{lammich:refinement-imperative}), but they target imperative rather than
functional programming, so we simply formalize the algorithm as an Isabelle
function. Isabelle/HOL provides an extensive library
of data structures and proofs; in this development we mainly rely on the set and
list libraries.
We start by introducing a \emph{locale}, fixing parameters and assumptions
for the remainder of the proof. We explicitly assume that the set of vertices is
finite: by default, sets may be infinite in Isabelle/HOL.

\clearpage
\begin{small}
\begin{lstlisting}[language=isabelle]
locale graph =
  fixes vertices :: 'v set
    and successors :: 'v => 'v set
  assumes finite vertices
      and forall v sin vertices. successors v subseteq vertices
\end{lstlisting}
\end{small}

\noindent%
We introduce reachability in graphs using an inductive predicate definition,
rather than via an explicit reference to paths as in the Why3 definition.
Isabelle then generates appropriate induction theorems for use in proofs.

\begin{small}
\begin{lstlisting}[language=isabelle]
inductive reachable where
  reachable x x
| [|y sin successors x; reachable y z|] ==> reachable x z
\end{lstlisting}
\end{small}

\noindent%
The definition of strongly connected components mirrors that used in Why3. The
following lemma states that SCCs are disjoint; its one-line proof is found
automatically using \emph{sledgehammer}~\cite{blanchette:sledgehammer}, which
heuristically selects suitable lemmas from the set of available facts (including
Isabelle's library), invokes several automatic provers, and in case of success
reconstructs a proof that is checked by the Isabelle kernel.

\begin{small}
\begin{lstlisting}[language=isabelle]
lemma scc-partition:
  assumes is-scc S and is-scc S' and x sin S cap S'
  shows   S = S'
\end{lstlisting}
\end{small}

\noindent%
Environments are represented by records, similar to the formalization in Why3,
except that there is no distinction between regular and ``ghost'' fields. Also,
the definition of the well-formedness predicate closely mirrors that used in
Why3.\footnote{We use infix syntax and the symbol $\preceq$ to denote precedence.
  The correspondence between numbers of vertices in the stack and precedence is
  asserted by the invariant \prog{wf\_num}.}

\begin{small}
\begin{lstlisting}[language=isabelle]
record 'v env =
  black :: 'v set            gray :: 'v set
  stack :: 'v list           sccs :: 'v set set
  sn :: nat                 num :: 'v => int

definition wf_env where wf_env e deq
  wf_color e /\ wf_num e
/\ distinct (stack e) /\ no_black_to_white e
/\ (forall x y. y prec x in (stack e) --> reachable x y)
/\ (forall y sin set (stack e). exists g sin gray e.
      y prec g in (stack e) /\ reachable y g)
/\ sccs e = { C . C subseteq black e /\ is_scc C }
\end{lstlisting}
\end{small}

\noindent%
The definition of the two mutually recursive functions \prog{dfs1} and
\prog{dfs} again closely follows their representation in Why3.

\begin{small}
\begin{lstlisting}[language=isabelle]
function (domintros) dfs1 and dfs where
  dfs1 x e =
   (let (n1,e1) = dfs (successors x) 
                      (add_stack_incr x e) in
     if n1 < int (sn e) then (n1, add_black x e1)
     else (let (l,r) = split_list x (stack e1) in
       (maxint,
        (| black = insert x (black e1),
          gray = gray e, stack = r, sn = sn e1,
          sccs = insert (set l) (sccs e1),
          num = set_infty l (num e1) |) ))) and
  dfs roots e =
   (if roots = {} then (maxint, e) 
    else (let x = SOME x. x sin roots;
              res1 = (if num e x <> -1
                      then (num e x, e)
                      else dfs1 x e);
              res2 = dfs (roots - {x}) (snd res1)
          in  (min (fst res1) (fst res2), 
                    snd res2) ))
\end{lstlisting}
\end{small}

\noindent%
The \textbf{function} keyword introduces the definition of a recursive function.
Isabelle checks that the definition is well-formed and generates appropriate
simplification and induction theorems. Because HOL is a logic of total
functions, it introduces two proof obligations: the first one requires the user
to prove that the cases in the function definitions cover all type-correct
arguments; this holds trivially for the above definitions. The second obligation
requires exhibiting a well-founded ordering on the function parameters that
ensures the termination of recursive function invocations, and Isabelle provides
a number of heuristics that work in many cases. However, the functions defined
above will in fact not terminate for arbitrary calls, in particular for
environments that assign sequence number $-1$ to non-white vertices. The
\prog{domintros} attribute instructs Isabelle to consider these functions as
``partial''. More precisely, it introduces an explicit predicate representing
the domains for which the functions are defined. This ``domain condition''
appears as a hypothesis in the simplification rules that mirror the function
definitions so that the user can assert the equality of the left- and right-hand
sides of the definitions only if the domain predicate holds. Isabelle also
proves (mutually inductive) rules for proving when the domain condition is
guaranteed to hold. Our first objective is therefore to establish sufficient
conditions that ensure the termination of the two functions. Assuming the domain
condition, we prove that the functions never decrease the set of colored
vertices and that vertices are never explicitly assigned the number $-1$ by our
functions. Denoting the union of gray and black vertices as \prog{colored}, we
define the predicate
\ednote{jjl}{-1 seems important for termination.. but the
  goal is to keep cardinal of white nodes positive, as white nodes are expressed
  by a difference on sets and cardinals are always non negative. No?

  sm: I'm afraid I don't understand what you mean. Perhaps we are talking at
  cross-purposes? I have slightly rephrased the above, is it clearer now?}

\begin{small}
\begin{lstlisting}[language=isabelle]
definition colored_num where colored_num e deq
  forall v sin colored e. v sin vertices /\ num e v <> -1
\end{lstlisting}
\end{small}

\noindent%
and show that this predicate is an invariant of the functions.
We can then show that the triple defined as

\begin{small}
\begin{lstlisting}[language=isabelle]
  (vertices - colored e, {x}, 1)
  (vertices - colored e, roots, 2)
\end{lstlisting}
\end{small}

\noindent%
for the arguments of \prog{dfs1} and \prog{dfs}, respectively, decreases w.r.t.\
lexicographical ordering on finite subset inclusion and $<$ on natural numbers
across recursive function calls, provided that \prog{colored\_num} holds when
the function is called and \prog{x} is a white vertex (resp., \prog{roots} is a
set of vertices). These conditions are therefore sufficient to ensure that the
domain condition holds:\footnote{Observe that Isabelle introduces a single
  operator corresponding to the two mutually recursive functions whose domain is
  the disjoint sum of the domains of both functions.}

\begin{small}
\begin{lstlisting}[language=isabelle]
theorem dfs1_dfs_termination:
  [|x sin vertices - colored e; colored_num e|] ==>
      dfs1_dfs_dom (Inl(x,e))
  [|roots subseteq vertices; colored_num e|] ==>
      dfs1_dfs_dom (Inr(roots,e))
\end{lstlisting}
\end{small}

\noindent%
The proof of partial correctness follows the same ideas as the proof presented
for Why3. We define the pre- and post-conditions of the two functions as
predicates in Isabelle. For example, the predicates for \prog{dfs1} are defined
as follows:

\begin{small}
\begin{lstlisting}[language=isabelle]
definition dfs1_pre where dfs1_pre e deq
    wf_env e /\ x sin vertices /\ x notin colored e
  /\ (forall g sin gray e. reachable g x)

definition dfs1_post where dfs1_post x e res deq
    let n = fst res; e' = snd res
    in  wf_env e' /\ subenv e e' /\ roots subseteq colored e'
      /\ (forall x sin roots. n <= num e' x)
      /\ (n = maxint \/ (exists x sin roots. exists y in set (stack e').
                      num e' y = n /\ reachable x y))
\end{lstlisting}
\end{small}

\noindent%
We now show the following theorems:
\begin{itemize}
\item The pre-condition of each function establishes the pre-condition of every
  recursive call appearing in the body of that function. For the second
  recursive call in the body of \prog{dfs} we also assume the post-condition of
  the first recursive call.
\item The pre-condition of each function, plus the post-con\-di\-tions of each
  recursive call in the body of that function, establishes the post-condition of
  the function.
\end{itemize}

\noindent%
Combining these results, we establish partial correctness:

\begin{small}
\begin{lstlisting}[language=isabelle]
theorem dfs_partial_correct:
  [|dfs1_dfs_dom (Inl(x,e)); dfs1_pre x e|] ==>
      dfs1_post x e (dfs1 x e)
  [|dfs1_dfs_dom (Inr(roots,e)); dfs_pre roots e|] ==>
      dfs_post roots e (dfs roots e)
\end{lstlisting}
\end{small}

\noindent%
We define the initial environment and the overall function.

\begin{small}
\begin{lstlisting}[language=isabelle]
definition init_env where init_env deq
  (| black = {},    gray = {},
    stack = [],    sccs = {},
    sn = 0,        num = lambda_. -1 |)
definition tarjan where tarjan deq
  sccs (snd (dfs vertices init_env))
\end{lstlisting}
\end{small}

\noindent%
It is trivial to show that the arguments to the call of \prog{dfs} in the
definition of \prog{tarjan} satisfy the pre-condition of \prog{dfs}. Putting
together the theorems establishing termination and partial correctness,
we obtain the desired total correctness results.

\begin{small}
\begin{lstlisting}[language=isabelle]
theorem dfs_correct:
  dfs1_pre x e ==> dfs1_post x e (dfs1 x e)
  dfs_pre roots e ==> dfs_post roots e (dfs roots e)
theorem tarjan_correct:
  tarjan = { C . is_scc C /\ C subseteq vertices }
\end{lstlisting}
\end{small}

\noindent%
The intermediate assertions appearing in the Why3 code guided the overall proof:
they are established either as separate lemmas or as intermediate steps within
the proofs of the above theorems. Similar as in the \Coq{} proof, the overall
induction proof was explicitly decomposed into individual lemmas as laid out
above. In particular, whereas Why3 identifies the predicates that can be used
from the function code and its annotation with pre- and post-conditions, these
assertions appear explicitly in the intermediate lemmas used in the proof of
theorem \prog{dfs\_partial\_correct}. The induction rules that Isabelle
generated from the function definitions was helpful for finding the appropriate
decomposition of the overall correctness proof.

Despite the extensive use of \emph{sledgehammer} for invoking automatic back-end
provers, including the SMT solvers CVC4 and Z3, from Isabelle, we found that in
comparison to Why3, significantly more user interactions were necessary in order
to guide the proof. Although many of those were straightforward, a few required
thinking about how a given assertion could be derived from the facts available
in the context. Table~\ref{table:isaline} indicates the distribution of the
number of interactions used for the proofs of the 46 lemmas the theory contains.
These numbers cannot be compared directly to those shown in
Table~\ref{table:coqline} for the \Coq{} proof because an Isabelle interaction
is typically much coarser-grained than a line in a \Coq{} proof. As in the case
of Why3 and \Coq{}, the proofs of partial correctness of \prog{dfs1} (split into
two lemmas following the case distinction) and of \prog{dfs} required the most
effort. It took about one person-month to carry out the case study, starting
from an initial version of the Why3 proof. Processing the entire Isabelle theory
on a laptop with a 2.7~GHz Intel\textsuperscript{\textregistered} Core i5
(dual-core) processor and 8~GB of RAM takes 35~seconds of CPU time.

\begin{table}[t]
\begin{center}
\begin{footnotesize}
\begin{tabular}{|c|c|c|c|c|c|c|c|}\hline
  $i = 1$ & $i \le 5$ & $i \le 10$ & $i \le 20$ & $i \le 30$ & $i = 35$ & $i = 43$ & $i = 48$ \\ \hline
    28    &   8       &    4       &    1       &   2        &    1     & 1        & 1\\  \hline
\end{tabular}
\end{footnotesize}
\end{center}
\caption{Distribution of interactions in the Isabelle proofs.}
\label{table:isaline}
\end{table}

\section{General comments about the proof}
\label{sec:proof-comments}

Our proof refers to colors, finite sets, and the stack, although the
general argument seems to be about properties of strongly connected
components in spanning trees. The algorithmician would explain the
algorithm with spanning trees as in Tarjan's article. It would be nice
to extract a program from such a proof, but beside the fact that this
is not so easy, programmers like to understand the proof in terms
of variables and data that their program is using.

A first version of the formal proof used \prog{ranks} in the working stack and a
flat representation of environments by adding extra arguments to functions for
the black, gray, sccs sets and the stack. That was perfect for the automatic provers
of Why3. But after remodelling the proof in Coq and Isabelle/HOL, it was simpler
to gather these extra arguments in records and have a single extra argument for
environments. Also \prog{ranks} disappeared leaving space to the \prog{num}
function and the precedence relation, which are easier to understand. The
automatic provers have more difficulties with the inlining of environments, but
with a few hints they could still succeed.

Finally, coloring of vertices is usual for graph algorithms, but we are
aware that a proof without colors is feasible and has indeed been done
without colors in Coq (see~\cite{CohenThery-SCC17}). The stack used in our
algorithm is also not necessary since it is just used to efficiently
output new strongly connected components. The proof can be implemented
with just finite sets, and the components will be obtained by computing
differences between visited sets of vertices. However, the stack-based
formulation ensures that the algorithm works in linear time,
and then it must be present in the proof, and its content must be related to the
visited sets of vertices.

There is thus a balance between the concision of the proof and
its relation to the real algorithm. In our presentation, we therefore
have allowed for a few redundancies.

\section{Conclusion}
\label{sec:conclusion}

The formal proof expressed in this article was initially designed and
implemented in Why3~\cite{ChenJJL17} after a long process, nearly a
2-year half-time work with many attempts of proofs about various graph
algorithms (depth first search, Kosaraju strong connectivity,
bi-connectivity, articulation points, minimum spanning tree). The big
advantage of Why3 is the clear separation between programs and the
logic with Hoare-logic style assertions and pre-/post-conditions about
functions. It makes the correctness proof quite readable for a
programmer. Also first-order logic is easy to understand. Moreover, one
can prove partial correctness without caring about termination.

Another important feature of Why3 is its interface with off-the-shelf
theorem provers (mainly SMT provers). Thus the system benefits of the
current technology in theorem provers and clerical sub-goals can be
delegated to these provers, which makes the overall proof shorter and
easier to understand. Although the proof must be split in more
elementary pieces, this has the benefit of improving its readability.
Several hints about inlining or induction reasoning are still
needed. Also, despite a useful XML file that records sessions and
facilitates incremental proofs, sometimes seemingly minor modifications to the
formulation of the algorithm or the predicates may result in the provers no
longer being able to handle a proof obligation automatically.

The \Coq{} and Isabelle proofs were inspired from the Why3 proof. Their
development therefore required much less time although their text is longer. The
\Coq{} proof uses the \textsc{SSReflect} macros and the Mathematical Components
library, which helps reduce the size of the proof compared to classical
\Coq{}. The proof also uses the bigops library and several other higher-order
features which makes the proof more abstract and closer to Tarjan's original
proof.

In \Coq{}, recursion cannot be used without proving termination. This
requires an agile treatment of mutually recursive definitions of
functions. Partial correctness can be proved by considering the
functionals of which the recursive definitions are the fixed point, and
passing as arguments the pre/post-conditions of these
functions. Moreover the recursive definitions take as extra argument
the number of recursive calls, in order to postpone the termination
argument.

Our \Coq{} proof does not use significant automation\footnote{Hammers exist for
  \Coq{}~\cite{CoqHammer,smtcoq} but unfortunately they currently perform badly
  when used in conjunction with the Mathematical Components library.}. All
details are explicitly expressed, but many of them were already present in the
Mathematical Components library. 
Moreover, a proof certificate is produced and a functional program could in
principle be extracted.
The absence of automation makes the system very
stable to use since the proof script is explicit, but it requires
a higher degree of expertise from the user. Still,
this lack of automation gives the user a direct feedback of how well 
the definitions work together.
This led us to develop an alternative and more concise (50\% shorter) formalization
without colors~\cite{CohenThery-SCC17}.

The Isabelle/HOL proof was the last one to be implemented. It closely follows
the Why3 proof, and can be seen as a mid-point between the Why3 and \Coq{}
proofs. It uses higher order logic and the level of abstraction is close to the
one of the \Coq{} proof, although more readable in this case study. The proof makes
use of Isabelle's extensive support for automation. In particular,
\emph{sledgehammer}~\cite{blanchette:sledgehammer} was very useful for finding
individual proof steps. It heuristically selects lemmas and facts available in
the context and then calls automatic provers (SMT solvers and superposition-based
provers for first-order logic). When one of these provers finds a proof,
sledgehammer attempts to find a proof that can be certified by the Isabelle
kernel, using various proof methods such as combinations of rewriting and
first-order reasoning (blast, fastforce etc.), calls to the \emph{metis} prover
or reconstruction of SMT proofs through the \emph{smt} proof method. Unlike in
Why3, the automatic provers used to find the initial proof are not part of the
trusted code base because ultimately the proof is checked by the kernel. The
price to pay is that the degree of automation in Isabelle is still significantly
lower compared to Why3. Adapting the proof to modified definitions was fast: the
Isabelle/jEdit GUI eagerly processes the proof script and quickly indicates
those steps that require attention.

The Isabelle proof also faces the termination problem to achieve general
consistency. Since termination cannot be ensured for arbitrary arguments, the
treatment of termination is delayed with the use of the \emph{domintros}
predicate. The proofs of termination and of partial correctness are independent; in
particular, we obtain a weaker predicate ensuring termination than the one used
for partial correctness. Although the basic principle of the termination proof
is very similar to the \Coq{} proof and relies on considering functionals of
which the recursive functions are fixpoints, the technical formulation is more
flexible because we rely on proving well-foundedness of an appropriate relation
rather than computing an explicit upper bound on the number of recursive calls.

One strong point of Isabelle/HOL is its nice \LaTeX{} output and the flexibility
of its parser, supporting mathematical symbols. Combined with the hierarchical
Isar proof language~\cite{wenzel:isar}, the proof is in principle
understandable without actually running the system, although some familiarity with
the system is still required.

In the end, the three systems Why3, \Coq{}, and Isabelle/HOL are mature, and
each one has its own advantages w.r.t.\ readability, expressivity, stability or
mechanization. Coming up with invariants that are both strong enough and
understandable was by far the hardest part in this work. This effort
requires creativity and understanding, although proof assistants provide some
help: missing predicates can be discovered by understanding which parts of the
proof fail. We think that formalizing the proof in all three systems was
very rewarding and helped us better understand the state of the art in
computer-aided deductive program verification. It could be also interesting to
experiment this proof in other formal systems and establish comparisons based on
this quite challenging example\footnote{
We have set up a Web page \url{http://www-sop.inria.fr/marelle/Tarjan/contributions.html}
in order to collect formalizations.}.

Another interesting work would be to verify an implementation of this
algorithm with imperative programs and concrete data structures. This
will complexify the proof, since mutable variables and mutable data
structures have to be considered. Several attempts were already
exposed~\cite{chargueraud-12-cf,chargueraud-16-horepr,lammich:refinement-imperative}
and it would be interesting to also develop them simultaneously in various
formal systems and to understand how these proofs can be derived from ours.

A final and totally different remark is about teaching of
algorithms. Do we want students to formally prove algorithms, or to
present algorithms with assertions, pre- and post-conditions, and make
them prove these assertions informally as exercises? In both cases,
we believe that our work could make a useful contribution.

\paragraph{Acknowledgements.}
We thank the Why3 team at Inria-Saclay/LRI-Orsay for very valuable
advice. This material is based upon work partly supported by the 
\texttt{proofinuse} project ANR-13-LAB3-0007.

\bibliography{scct}



\end{document}